\documentclass[natbib,smallextended]{svjour3}

\usepackage{graphicx}
\usepackage{hyperref}
\usepackage{url}
\usepackage{comment}
\usepackage{color}
\usepackage{multirow}
\usepackage[most]{tcolorbox}
\usepackage{tabularx}
\usepackage{balance}
\usepackage{changepage}

\journalname{Empirical Software Engineering}

\begin{document}

\title{Developers Perception of Peer Code Review in Research Software Development}

\author{Nasir U. Eisty         \and
        Jeffrey C. Carver %etc.
}

\institute{N. U. Eisty \at
            Department of Computer Science \\
            Boise State University \\
            Boise, ID, USA \\
            \email{nasireisty@boisestate.edu}
           \and
           J. C. Carver \at
              Department of Computer Science\\
              University of Alabama\\
             Tuscaloosa, AL, USA\\
              \email{carver@cs.ua.edu}
}

\date{Received: date / Accepted: date}

\maketitle

\begin{abstract}
\textit{Background}: 
Research software is software developed by and/or used by researchers, across a wide variety of domains, to perform their research.
Because of the complexity of research software, developers cannot conduct exhaustive testing.
As a result, researchers have lower confidence in the correctness of the output of the software.
Peer code review, a standard software engineering practice, has helped address this problem in other types of software.
\textit{Aims}: 
Peer code review is less prevalent in research software than it is in other types of software.
In addition, the literature does not contain any studies about the use of peer code review in research software.
Therefore, through analyzing developers perceptions, the goal of this work is to understand the current practice of peer code review in the development of research software, identify challenges and barriers associated with peer code review in research software, and present approaches to improve the peer code review in research software.
\textit{Method}: 
We conducted interviews and a community survey of research software developers to collect information about their current peer code review practices, difficulties they face, and how they address those difficulties.
\textit{Results}: 
We received 84 unique responses from the interviews and surveys.
The results show that while research software teams review a large amount of their code, they lack formal process, proper organization, and adequate people to perform the reviews. 
\textit{Conclusions}: 
Use of peer code review is promising for improving the quality of research software and thereby improving the trustworthiness of the underlying research results.
In addition, by using peer code review, research software developers produce more readable and understandable code, which will be easier to maintain.

\end{abstract}

\section{Introduction} 
\label{sec:introduction}
Researchers in a number of scientific, engineering, business, and humanities domains increasingly develop and/or use software to conduct or support their research~\citep{5069155}.
We refer to this software collectively as \textit{research software}.
Without such research software, it would be difficult or impossible for many researchers to do their work~\citep{nangia_katz_2017,5069155}.
Research software includes both software for end-user researchers, such as for weather forecasting or molecular dynamics simulation, and software that provides infrastructure support, including messaging middleware, scheduling software, and various mathematical, scientific, or statistical libraries.
Results produced by research software help researchers predict natural phenomena and make decisions to support critical needs.
Researchers need high quality software that will produce trustworthy results and function properly in mission-critical situations.
Therefore, researchers need to follow appropriate software engineering practices to help ensure the quality of the software.

However, in practice, the development of research software differs significantly from the development of more traditional commercial/IT software as discussed below~\citep[Introduction]{1958562}.
There are a number of factors that result in this difference.
First, research software has additional risks due to the exploration of relatively unknown scientific phenomenon and the essential complexity of the research domain.
Second, the requirements for research software change constantly as researchers gather new information from results of simulation or new discoveries.
Third, research software often has complex communication or I/O patterns along with data dependency that can degrade performance.
Fourth, research software developers often need highly specialized skills related to numerical algorithms and systems. 
Fifth, the larger node/core sizes and physical computing facilities required by research software along with the longer runtimes result in an increased likelihood of computing failures.
As a result of these factors, the software engineering tools typically used by traditional software developers often will not work in the computing environment or for the typical developers of research software.

Due to the nature of research software, research software developers often emphasize their scientific goals over software quality and maintainability goals~\citep{article_segal} and over the use of appropriate software engineering practices~\citep{10.1145/1370175.1370252}.
One potential reason why research software developers have different goals that developers of commercial/IT or open-source software is that the knowledge, skills, and incentives present in research software development differ from those present in traditional software domains~\citep{4548404}.
For example, research software developers often lack formal software engineering training while trained software engineers may lack the required depth of understanding of the underlying research domain. 
In addition, the incentives in the research domain focus of timely results and publications, rather than more traditional software quality/productivity goals.
There have been many efforts to understand how software engineering can help with the development and maintenance of research software.
However, there is little evidence that these efforts have led research software teams to value software engineering practices in a manner similar to how more traditional software teams value software engineering practices~\citep{5337642}.

Historically, developers of business/IT software have employed various types of testing techniques to improve the quality of their software.
While testing is a useful practice, there are some technical challenges for testing research software.
The first challenge is the lack of test oracles~\citep{5069163}.
An oracle is pragmatically unattainable in most of the cases for research software because researchers develop software to find previously unknown answers.
Due to the lack of test oracles, research software developers often use judgment and experience to check the correctness of the software. 
The second challenge is the large number of tests required to test research software using standard testing techniques.
Also, the large number of input parameters makes it challenging to manually selecting a sufficient test suite~\citep{10.1007/978-3-540-69389-5_34}.
Finally, the presence of legacy code makes testing research software  challenging~\citep{5999647}.
However, developers of research software have found it more difficult to employ some of the traditional software testing techniques~\citep{KANEWALA20141219}.

The complexity of the underlying research domains leads to research software that has complex computational behavior.
This complexity, along with the fact that the expected outputs are often unknown, make it difficult to define appropriate tests and identify input domain boundaries~\citep{6086527}.
In many cases, the input space of research software is so vast that it is not feasible, or even possible, for a developer to create a test suite that adequately exercises the limits of the software~\citep{10.1007/978-3-540-69389-5_34}.
Therefore, while testing is useful, it is generally not sufficient to ensure the quality of research software.

Conversely, peer code review is a lightweight, asynchronous method for ensuring high-quality code~\citep{28121}.
Peer code review is a systematic examination of source code by peers of the software's developer to identify problems the developer can then address.
As defined in the literature~\citep{6681346}, peer code review is the process of analyzing code written by a teammate (i.e. a peer) to judge whether it is of sufficient quality to be integrated into the main project codebase.
Traditionally, commercial organizations and open source projects have been adopting peer code review as a more efficient, lightweight version of the older, more formal inspection process~\citep{6148202}.
While peer code review is effective and prevalent in open-source and commercial software projects, it remains underutilized in research software~\citep{article159}.

In addition to improving general software quality, the use of peer code review has other specific benefits in the research software domain.
Unlike traditional commercial/IT software, research software developers are often exploring new scientific or engineering results, which may be unknown \textit{a priori}. 
The lack of an oracle makes it difficult for developers to create adequate tests that can diagnose whether a result is a new insight from a simulation or is the consequence of a fault in the software~\citep{HEATON2015207}.
Even in cases where the expected output is known, the complexity of the software often makes it impossible to adequately test all important configurations of the software and input data.
Conversely, when a person conducts a code review, he or she is able to analyze the underlying algorithm and identify problematic conditions.
Therefore, while peer code review is essential for any type of software, it is even more important for research software~\citep{article159}.

By employing peer code review in their projects, research software developers will see benefits both in the short term, through higher quality scientific results produced by high-quality software, and in the long term, through creation of more maintainable software~\citep{10.1145/2597073.2597076}.
The higher quality scientific results occur because developers focus their attention on the code itself to identify mistakes, inefficiencies, and other aspects of the code that need improvement.
The improved maintainability arises because as team members start reviewing each other's code, they begin writing more readable code to enable the peer-review process. 
Code that is more readable and easier to understand is also more maintainable over time.

However, despite the benefits peer code review can provide, developers of research software do not yet apply it widely~\citep{article159}.
In addition, the literature does not contain any studies on the use of peer code review in research software.
Therefore, to better understand how to increase the usage of peer code review in research software development, this paper focuses on characterizing how research software developers perform peer code review.
The primary goal of this study is to \textit{better understand the practices, difficulties, impacts, and potential areas of improvement of peer code review in research software development}. 
To gather this insight, we interviewed and surveyed research software developers from a wide array of projects about their current practice and the challenges they face.

The key contributions of this paper are:
\begin{itemize}
	\item The first study about the use of peer code review in the development of research software.
	\item An overview of the current peer code review practices in research software development.
	\item Identification of the types of defects identified during the peer code review process.
	\item Positive and negative experiences research software developers have regarding peer code review.
	\item Impacts of the peer code review process on research software projects. 
	\item Challenges and barriers developers face during peer code review.
	\item Potential areas of improvement in the peer code review process.
\end{itemize}

The remainder of this paper is organized as follows.
In Section~\ref{sec:literature} we provide the necessary background information to motivate the research questions explored in this study.
In Section~\ref{sec:Methodology} we explain the study design, data collection, and data analysis procedures.
In Section~\ref{sec:Results} we provide the detailed results.
In Section~\ref{sec:Discussion} we discuss the key finding from study.
In Section~\ref{sec:Threats} we discuss the validity threats.
In Section~\ref{sec:Conclusion} we draw conclusions.

\section{Background}
\label{sec:literature}
Because peer code review has not yet been studied in the context of research software, this section focuses on prior research about peer code review in general.
The idea of a formal inspection first appeared in 1976~\citep{5388086}.
As initially developed, the process was very structured and heavyweight.
Reports of the use of this type of a formal inspection process showed that it improved product quality, productivity, and manageability~\citep{28121}.
However, as software development became more iterative and agile, such a heavyweight process was seen as less relevant.
Therefore, the formal inspection process has changed into a more lightweight, peer-review process.
Developers of open-source and commercial software have increasingly employed this more lightweight peer code review as part of their development process~\citep{10.1145/2597073.2597082, 1241366}.

Google began using peer code review early in its history and evolved the practice over the years. 
It is an important aspect of the development workflow and has become a core part of Google culture.
Because code acts as a teacher to future developers, Google forces people to write code that is understandable to other developers. 
Google developers expect four key themes from peer code review: education, maintaining norms, gatekeeping, and accident prevention.
Google does peer code review not only to solve problems but to ensure readability and maintainability \citep{Sadowski:2018:MCR:3183519.3183525,Potvin:2016:WGS:2963119.2854146}.

Open source software projects have also found peer code review a valuable method to improve software quality.
Components that have higher review coverage and higher rates of participation in peer code review are likely to have fewer post-release defects \citep{7081827}. 
Additionally, 75\% of the changes resulting from peer code review relate to extra-functional improvements, i.e. code quality and documentation.
Only 25\% relate to functionality \citep{Beller:2014:MCR:2597073.2597082,Bacchelli:2013:EOC:2486788.2486882,Mantyla:2009:TDR:1591905.1592371}.
That observation indicates that peer code review is valuable for short term defect detection but more valuable in long term maintainability.
Overall, poorly reviewed code has a negative impact on software quality in open source projects \citep{7081827}.

In a study comparing peer code review practices at Microsoft and open source projects, researchers studied the effects of peer code review on developers perception, collaboration, and other non-technical aspects of peer code review.
When it comes to accepting a review request, reviewers consider their relationship with the author, the reputation of the author, their own area(s) of expertise, and the anticipated time/effort required to conduct the review.
After accepting a code review request, reviewers typically start with the most familiar portions of the code before reviewing the unfamiliar parts, in which they often learn something.
Besides learning, peer code reviews ensure that at least one or two people other than the developer are aware of code changes, thus creating project awareness among the team members.
Furthermore, peer code reviews encourage community building and collaboration by fostering direct interactions between developers and reviewers \citep{7484733}.

Open source contributors come from diverse backgrounds, expertise levels, and locations. 
Because the quality of their contributed code varies greatly, the peer code reviews in open source projects emphasize maintainability and consistency.
Conversely, developers from Microsoft have less variation in their code.
Therefore, the peer code reviews can focus more on defect detection.
Microsoft developers consider peer code review as a great source of knowledge sharing. 
On average a Microsoft developer spends 15\%-25\% of his time in peer code review while an open source developer spends even more time on review \citep{7081827}.

Peer code reviews should occur early and often to help developers find problems before they become more expensive to fix.
It is easier for reviewers to focus on small, independent, and complete contributions.
But breaking up a large change or segment of code to make small pieces may cause new problems unless a divide-and-conquer approach is applied \citep{Rigby:2008:OSS:1368088.1368162}. 
Google keeps their peer code review process lightweight and flexible.
Over time, Google has converted their process into one with markedly smaller changes and quicker reviews \citep{Sadowski:2018:MCR:3183519.3183525}.

While there is extensive research on peer code review in commercial and open source software, there is no empirical research on the use of peer code review in research software.
Therefore, we must first gain a basic understanding of the process research software developers follow when performing peer code review.
We pose the following research question to gather developers perceptions about the peer code review process research software developers follow along with their positive and negative experiences.

\vspace{8pt}
\begin{tcolorbox}
\textbf{RQ1: How do research software developers perform peer code review?}
\end{tcolorbox}

One of the effects of peer code review is that it helps developers form impressions of their teammates.
The quality of the submitted code is key in this impression formation process. 
For instance, a developer who submits simple, understandable, and self-documenting code that requires minimum review time may gain improved social status.
Another effect of peer code review is participation in the peer code review process has strong positive effects on future collaborations.
As a code author's reputation increases based on high-quality code submissions, it helps her future code submissions.
In addition, as an author's reputation increases, people begin to believe more in her expertise and request her to review their code.
A reviewer's perception of the expertise of the code author also influences the level of scrutiny she uses during the peer code review~\citep{6681346,7484733}.

Based on these earlier findings, we pose the following question to understand whether the benefits of peer code review seen in open source and commercial software (discussed above) also appear in research software.
To understand the prevalence of these benefits, we need to gather developers perceptions about how peer code review impacts research software.

\vspace{4pt}
\begin{tcolorbox}
\textbf{RQ2: What effect does peer code review have on research software?}
\end{tcolorbox}

Developers and reviewers also face challenges in performing peer code review.
Reviewers find that understanding the code and the motivation for a change are the most difficult aspects of peer code review.
In addition, many people think peer code review is time-consuming, with the task of understanding the change being the most time-consuming \citep{5070996}.
Despite being a time consuming and challenging process, developers want to participate in peer code review because it helps with consistency, reliability, and maintainability of the project \citep{Bacchelli:2013:EOC:2486788.2486882}.
Google, Microsoft, Facebook, and many other organizations have improved their peer code review process through building tools, overcoming social challenges, and investing in the process \citep{Sadowski:2018:MCR:3183519.3183525,7180075}

The following research question seeks to understand developers perceptions whether those difficulties also hold for research software.
We need to collect information about the challenges and barriers research software developers face.

\vspace{4pt}
\begin{tcolorbox}
\textbf{RQ3: What difficulties do research software developers face with peer code review?}
\end{tcolorbox}

Similar to the improvements to tooling, processes, and investments commercial and open-source projects have made relative to the peer code review process, this question seeks to understand changes that could help peer code review for research software.
To answer the following research question we need to developers perceptions about the gaps in the process and the desired improvements.

\vspace{4pt}
\begin{tcolorbox}
\textbf{RQ4: What improvements to the peer code review process do research software developers need?}
\end{tcolorbox}

\section{Methodology}
\label{sec:Methodology}
For this study, we gathered data from two sources: interviews and surveys.
Most of the questions in this study focus on the developers' perception of peer code review.
Therefore, to obtain this type of information, we decided to ask developers directly rather than use repository mining.
While repository mining can provide interesting insights based on the information captured in in repositories, gathering information directly from people provides the types of explanations that are not available from repository mining alone.
This section describes the survey/interview questions, the methods for collecting data, and the data analysis process.

\subsection{Survey/Interview Design}
We designed the survey around the research questions defined in Section~\ref{sec:literature}. 
The first author reviewed the literature~\citep{6681346,7484733,BOSU20144} of similar types of studies and developed a list of potential interview and survey questions related to each research questions. 
Then both authors discussed this list of questions, kept Q3, Q4, Q5, Q7, Q11, and Q17, and formulated the rest of the questions in~\ref{fig_survey_questions} to collect adequate information to answer the research questions.

We piloted the initial set of questions with four research software developers to ensure the questions were clearly understandable.
Based on their feedback, we made the following updates to the questions:
(1) rephrased some questions to make them more easily understandable by research software developers,
(2) removed software engineering terminology that may be unfamiliar to potential respondents, and
(3) rearranged the question flow to be more logical.

Figure~\ref{fig_survey_questions} contains the final questions (after updates from the pilot) organized by research question.
The primary difference between how we asked these questions in the interviews and in the surveys is in the interviews, we asked all questions as open-ended questions (except for Q5).
We then derived the multiple choice answers listed after each question in Figure~\ref{fig_survey_questions} based on the responses to the survey.
In addition to asking questions related to the four research questions, we also included general questions to describe the demographics of the respondents.  
These questions help characterize the sample to provide additional confidence in the results.

\subsection{Data Collection}
First, we conducted interviews of research software developers.
Then, to reach a broader audience, we conducted a survey.
The following subsections detail each data collection method.

\subsubsection{Interviews}
The first author conducted the interviews while at NCSA for a summer internship.
We used convenience sampling to recruit 22 interviewees from two sources. 
First, 13 interviewees were NCSA research software developers drawn from three projects in different domains.
Second, 9 interviewees were attendees of the annual Einstein toolkit meeting\footnote{http://www.ncsa.illinois.edu/Conferences/ETK17/} who worked on projects from a wide variety of research domains including Physics, Astronomy, High Performance Computing, and Climate Science.
The interviewer used the questions in Figure~\ref{fig_survey_questions} as an interview guide, but adapted based on the responses of the interviewee, i.e. reordering questions as needed.

\subsubsection{Survey}
After completing the interviews, we then encoded the questions shown in Figure~\ref{fig_survey_questions} into the Qualtrics survey tool\footnote{https://www.qualtrics.com/}.
We started by targeting Computational Science and Engineering (CSE) projects.
Then, we broadened the scope to include a broader audience of research software developers so we could gather input from a diverse set of respondents.

To reach a broad sample within the target population, we employed a number of solicitation methods.
First, we sent the survey to contributors from the projects represented by the interviewees (but excluded the interviewees).
Second, we advertised the survey at the 2017 International Workshop on Software Engineering for High Performance Computing in Computational and Data-Enabled Science and Engineering\footnote{https://se4science.github.io/SE-CODESE17/}.
Third, a collaborator sent the survey to a mailing list of research software developers in the UK.
Fourth, a survey respondent shared the survey on the Research Software Engineers Slack channel.
Fifth, we advertised the survey in the monthly newsletter of the Better Scientific Software (http://bssw.io) organization.
Finally, we asked respondents to forward the survey invitation within their own networks.
As a result of this solicitation approach, we cannot estimate the number of people who received the invitation.

\begin{figure*}[p]
\caption{Survey Questions}
\label{fig_survey_questions}
\input{fig_surveyquestions}
\end{figure*}

\subsection{Data Analysis}
Prior to performing the data analysis, we examined the responses to ensure we included only valid ones.
Due to the length of the survey, some respondents did not answer all questions.
We defined a valid response as one that answered all quantitative questions and at least one qualitative question.
The first author excluded any responses that did not meet this criteria.
Then, the first author transcribed the audio recordings from the 22 interviews.
Because we performed the interviews before the survey, we asked the interviewees to refrain from taking the survey.
We did not want to bias the results by having two sets of answers from the same person.
Therefore, there is no overlap between the interview and the survey participants.
Together the interviews and surveys produced data from 84 unique participants for the analysis.

We used SPSS to compute frequency distributions for the quantitative data. 
We used NVivo to analyze the qualitative data.
We use R to visualize the data.
We used a standard qualitative analysis approach to code the survey and interview data, as follows.
First, each author individually coded the qualitative responses with one or more codes.
Then, we compared the results of the individual coding activities, consolidated items that had similar codes, and identified items where we disagreed.
We then discussed and resolved each disagreement.
Examples of the most common types of disagreements include:
(1) items where we code a response with semantically different codes, which we resolved by discussing and agreeing on the most suitable code (or codes),
(2) items where we code a response with differently worded but semantically similar codes, which we resolved by choosing one of the terms.
In the end, we resolved all disagreements and arrive at a final agreed-upon coding result.
We counted the total number of occurrences to calculate the frequency of each code and visualize the codes accordingly to charts.

\section{Results}
\label{sec:Results}
We organize this section around the four research questions.
Before discussing the specific results, we first provide an overview of the demographics to characterize the sample.
Throughout this section, the question numbers refer to the survey questions in Figure~\ref{fig_survey_questions}.
For the free response questions, our analysis could assign multiple codes to an individual answer. 
Therefore, in many of the analyses below, the sum of the responses is larger than the number of participants.
In the spirit of Open Science, we have made our de-identified raw data available~\citep{dataset}.

\subsection{\textit{\textbf{Demographics}}}
The answers to Q1 showed participants represent at least 45 research software projects, 13 participants chose not to reveal their project name for privacy reasons.
The answers to Q4 indicated that 72\% of respondents received some sort of financial compensation for participating in their projects.
The answer to these questions indicates that the participants come from a wide variety of projects and have a strong tie to those projects because of the financial compensation.

\begin{figure*}[!htb]
	\includegraphics[width=0.75\textwidth]{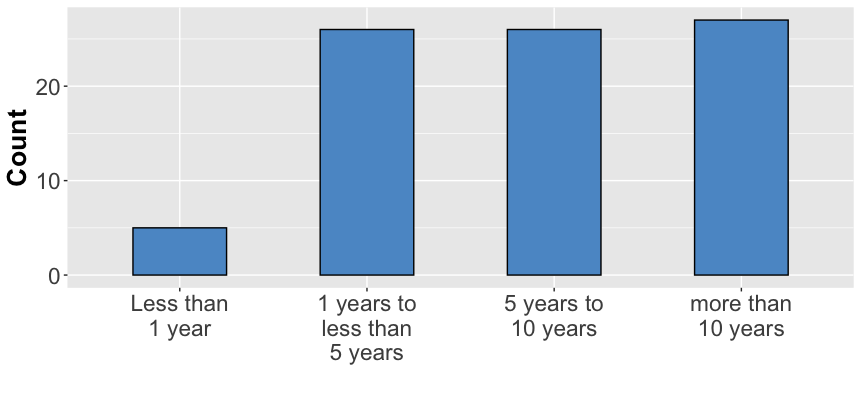}
	\caption{Number of years worked on research software}
	\label{fig_years_worked}
\end{figure*}

The distribution of responses to Q2 (Figure~\ref{fig_years_worked}), indicates most participants had at least 5 years of experience working in research software. 
Just under $1/3$ had more than ten years experience.
Only a small number had less than one year.
This distribution suggests that the participants had appropriate experience and knowledge to provide valid answers to the questions.

\begin{figure*}[!htb]
	\includegraphics[width=0.75\textwidth]{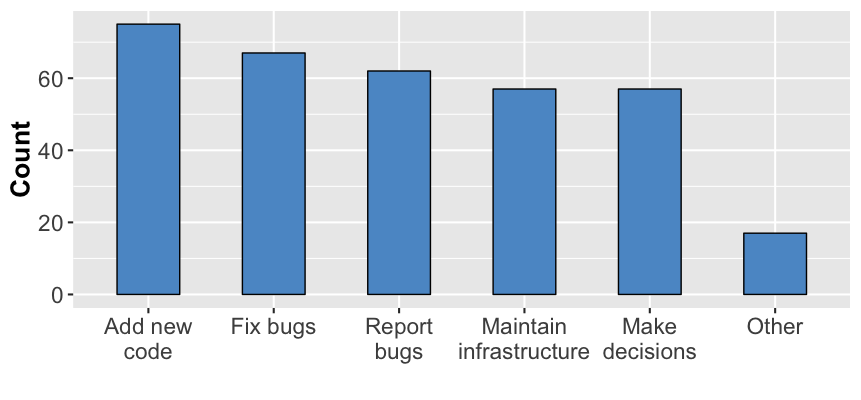}
	\caption{Respondents' role on project}
	\label{fig_role_on_project}
\end{figure*}

The distribution of responses to Q3 (Figure~\ref{fig_role_on_project}), indicates that the study participants assume different roles within their respective projects.

\begin{figure*}[!htb]
	\includegraphics[width=0.75\textwidth]{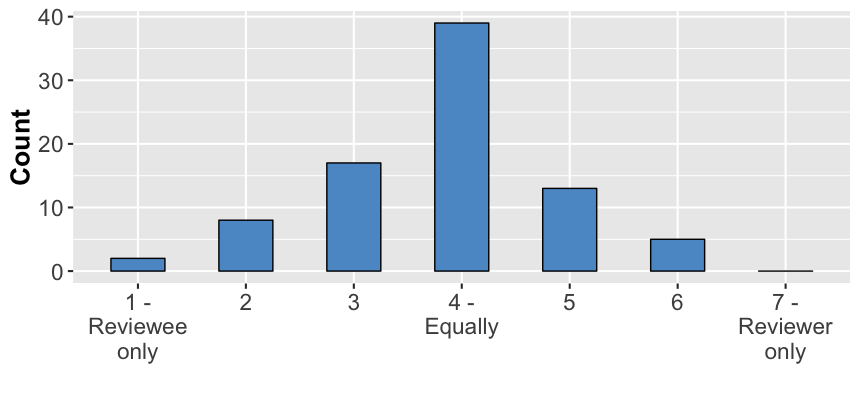}
	\caption{Balance as a reviewee and reviewer}
	\label{fig_balance_reviewee_reviewer}
\end{figure*}

Finally, the answers to Q5 (Figure~\ref{fig_balance_reviewee_reviewer}) show the respondents overwhelmingly participate both as a code reviewer and reviewee.
Only a small number of respondents act exclusively as either a reviewer or as a reviewee.
%An analysis of the raw data indicated that participants with more experience tend to review more code, while those with less experience tend to write more code.
Therefore, the study participants have appropriate expertise both with reviewing code and with receiving feedback from reviews to provide valuable insights into the peer code review process.

\subsection{\textit{\textbf{RQ1: How do research software developers perform peer code review?}}}
The following text discusses the overall practices of peer code review along with the respondents' experiences associated with the peer code review process.

\vspace{4pt}
\noindent
\textbf{Overall practices}

\noindent
In response to Q6, respondents described their peer code review process.
Overall, the respondents' projects typically follow an informal peer code review process. 
Sixty-one respondents indicated they initiate peer code review with their peers through pull-request on GitHub, Bitbucket, or GitLab. 
Respondents from larger projects combine an internal ticketing system with the pull requests. 
The responses varied as to how many people had to review each change or pull-request (anywhere from one to three) before merging into the main branch.
In some cases, small changes or bug fixes from experienced or core developers could bypass the review process entirely.

\begin{figure*}[!htb]
	\includegraphics[width=0.75\textwidth]{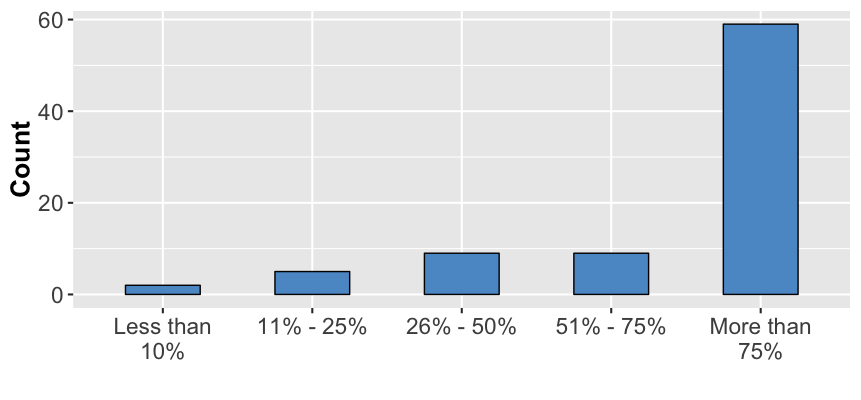}
	\caption{Percentage of code undergo review}
	\label{fig_code_undergo_review}
\end{figure*}

The responses to Q7 (Figure~\ref{fig_code_undergo_review}) show that, in the projects represented by the respondents, more than 75\% of the code undergoes peer review. 
This result is consistent with the previous research on the percentage of code typically reviewed in commercial and open source software~\citep{7484733}.

\begin{figure*}[!htb]
	\includegraphics[width=0.75\textwidth]{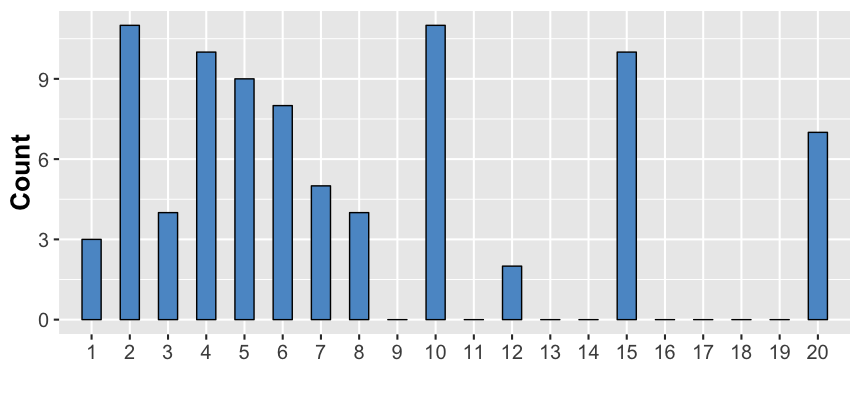}
	\caption{Number of reviewers}
	\label{fig_number_of_reviewer}
\end{figure*}

The responses to Q8 (Figure~\ref{fig_number_of_reviewer}) shows a wide variety in the number of people that participate in code review.
A further analysis of the raw data suggests that in the larger open-source research projects, only core developers perform code review, while in the smaller projects, almost all of the developers perform peer code review.
This observation makes sense as participants in smaller projects have to take on more tasks.

\vspace{4pt}
\noindent
\textbf{Acceptance of review requests}

\begin{figure*}[!htb]
	\includegraphics[width=0.75\textwidth]{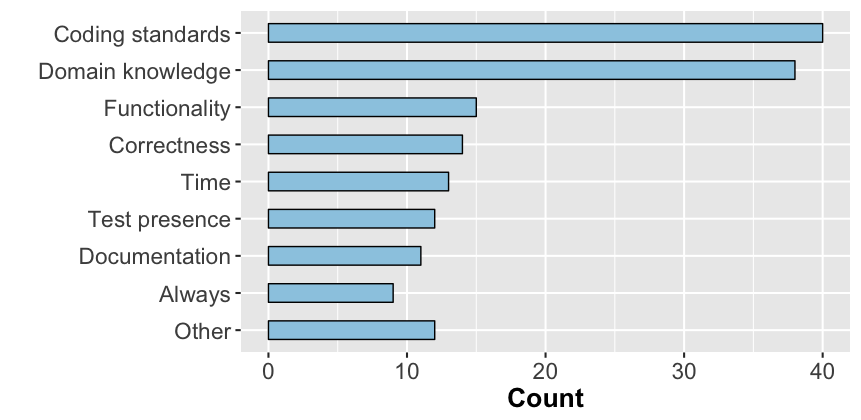}
	\caption{Factors to accept peer code review request}
	\label{fig_factors_to_accept_review_request}
\end{figure*}

\noindent
The results from Q9 (Figure~\ref{fig_factors_to_accept_review_request}) show which factors affect the participants decision to accept a peer code review request.
The most common factor is that the code follows \textbf{coding standards}.
As one respondent stated, \textit{``all changes need proper style (PEP-8 compliant at minimum) and need to pass the test-suite. Bug fixes just need to fix the known issue and not break any APIs, and potentially add relevant new tests. New features need to be fully documented and have tests.''}

The second most common factor in deciding whether to accept a review request is \textbf{domain knowledge}.
Reviewers want to \textit{``...have [domain] knowledge of the project, of the language, or the intended functionality.''}
They also want to know \textit{``the relevance of the review and whether or not, I am qualified to do the review''}.
The belief that \textit{``if i am suitable for the context of the change''} summarizes the role of domain knowledge in the decision to accept a review request.

Other factors that influence the decision to accept a review request were \textbf{correctness} of the code and whether its \textbf{functionality} addresses the project needs.
The answers that fell into the ``other'' category are generally not interesting. 
In most cases, only one or two responses were similar enough for us to group them, which is why we collected them all into the ``other'' category rather than listing them separately.
For example, three participants mentioned \textit{other potential reviews}, another 2 participants indicated \textit{admin approval}, and 1 participant said \textit{politeness of request}. 

In addition, some participants always accept a peer code review request.
For any requests that are outside their expertise, these participants then refer the review to someone more appropriate.

\vspace{4pt}
\noindent
\textbf{Amount of Code Reviewed}

\noindent
The answers to Q10 varied from project to project.
Because the survey question did not specify the unit of measure, the respondents took three perspectives on describing how much code they reviewed at one time.

The first perspective, mentioned by 37 respondents, used files as the unit of measure.
One participant said the amount of code reviewed \textit{``varies greatly, from 1 file to 80 files. Most reviews involve about 5-20 files.''}
These files usually contain a subroutine, module, or a small code segment but \textit{``this depends on the size of the merge request, though we encourage small merge requests to make the job manageable.''}

The second perspective, mentioned by 21 respondents, used lines of code (LOC) as the unit of measure. 
One said \textit{``this [the size of the reviewed change] varies widely. Student PRs are often less than 100 lines, but need many iterations. Developer commits are sometimes much longer, but need less work.''}
Typically it is \textit{``50-200 lines of code plus associated tests/documentation ($\sim$500 lines total), but occasionally ranging from 5 to 10,000 lines.''}

The third perspective, mentioned by 26 participants, used pull-request (PRs) as the unit of measure.
\textit{``[A] PR is usually one physics module, new problem type, etc.  It's generally around a few hundred lines of code at a time, max.''}
Participants review from \textit{``one PR, preferably smaller than 1000 LOC or at least broken into smaller commits, but sometimes large changes''} to a few PR per day.

\vspace{4pt}
\noindent
\textbf{Time Spent Reviewing Code}

\begin{figure*}[!htb]
	\includegraphics[width=0.75\textwidth]{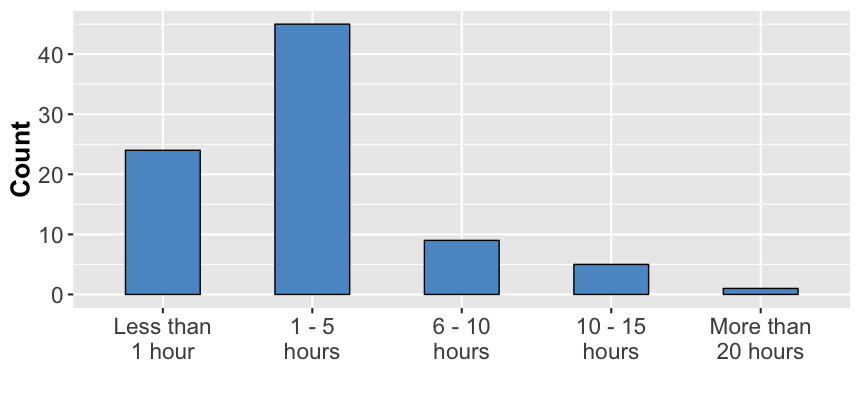}
	\caption{Time spent on peer code review}
	\label{fig_spend_time_on_review}
\end{figure*}

\noindent
The responses to Q11 (Figure~\ref{fig_spend_time_on_review}) show that most respondents spend one to five hours per week on peer code review.
An additional $1/3$ of respondents spend less than one hour per week.
Still fewer respondents spend more than five hours per week.
This result is similar to previous results about the amount of time spent in peer code review for commercial and open source software projects \citep{7484733}.

\begin{figure*}[!htb]
	\includegraphics[width=0.75\textwidth]{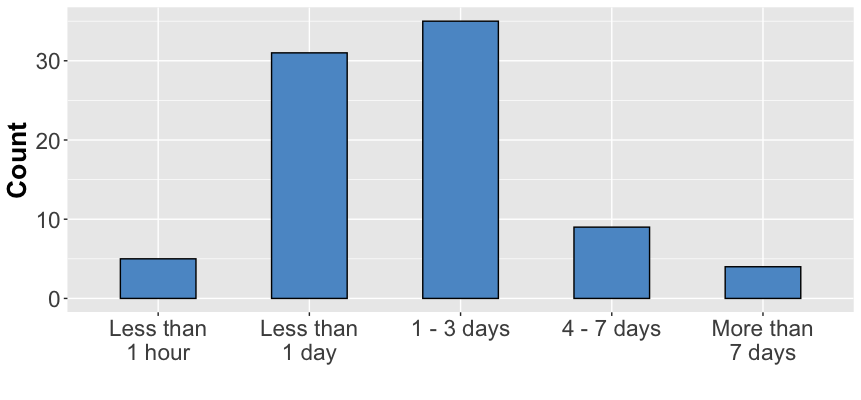}
	\caption{Time for a first response}
	\label{fig_first_response_time}
\end{figure*}

The responses to Q12 (Figure~\ref{fig_first_response_time}) show how long it takes a code author to get the first feedback on his or her code submission.
Interestingly, 43\% of the respondents said it takes less than a day for a first response.
This result is surprising given that conducting peer code review is not the primary job of research software developers, nor do they receive incentives for performing peer code review.
Approximately 40\% of the respondents indicated the response time was 1-3 days.
This result seems reasonable given that research software developers have other tasks related to their own work and may not prioritize reviewing other's code quite as highly.

\begin{figure*}[!htb]
	\includegraphics[width=0.75\textwidth]{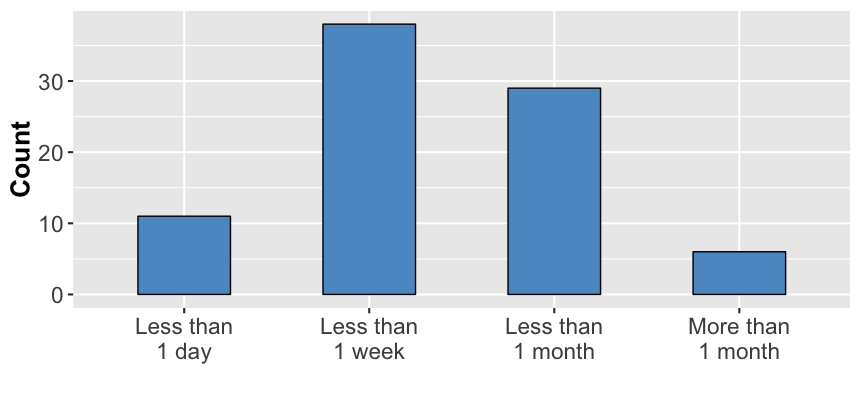}
	\caption{Time for a final decision}
	\label{fig_final_response_time}
\end{figure*}

The responses to Q13 (Figure~\ref{fig_final_response_time}) provide insight into the overall amount of time taken to reach a final decision on a code submission.
Approximately 45\% of respondents indicated the project reached a final decision in less than a week.
For very small changes or bug fixes, that time was even shorter (less than a day).
Overall, 93\% of the respondents indicated projects reach a decision in less than a month. 
The results from the interviews provided some insight into the time required for a final decision.

According to one, ``Depends on the size of the contributions--small things could be accepted in a day or two. Large things could take weeks or even months.''
The interviewees indicated that longer review times were often not the result of a review request sitting idle.
In cases where the change is larger or represents a new feature, it may take longer for reviewers to provide proper feedback.
According to another, ``This [the review time] is incredibly variable, and depends on the reviewers, what changes are requested, and the time it takes for the person issuing the PR to answer. I'd say that the median time is a week, with the mean being higher because some PRs take weeks to months due to slow response times.''

\vspace{4pt}
\noindent
\textbf{Problems Identified}

\begin{figure*}[!htb]
	\includegraphics[width=0.75\textwidth]{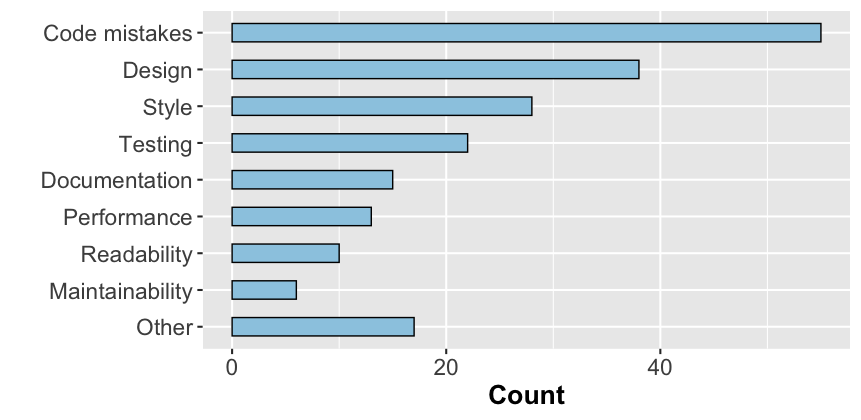}
	\caption{Problems peer code review identifies}
	\label{fig_problems_identified}
\end{figure*}

\noindent
The responses to Q14 indicated peer code reviews help research software developers identify many problems in the code.
Figure~\ref{fig_problems_identified} shows the types of problems reviewers identify.
Most of the respondents identified various type of problems related to software quality, including \textbf{code mistakes}, \textbf{design}, \textbf{style}, and \textbf{testing}.
As one respondents mentioned peer code review identifies \textit{``all sorts of stuff. Poor code, actual bugs, poor usage of macros, code with too narrowly-defined purpose that could easily be generalized.''}
Similarly, peer code reviews \textit{``identify corner cases that the author may not have seen, performance issues that might be introduced, breakages with other pieces of the software, algorithmic problems, typos, lack of comments/documentation.''}

\vspace{4pt}
\noindent
\textbf{Positive peer code review experiences}

\begin{figure*}[!htb]
	\includegraphics[width=0.75\textwidth]{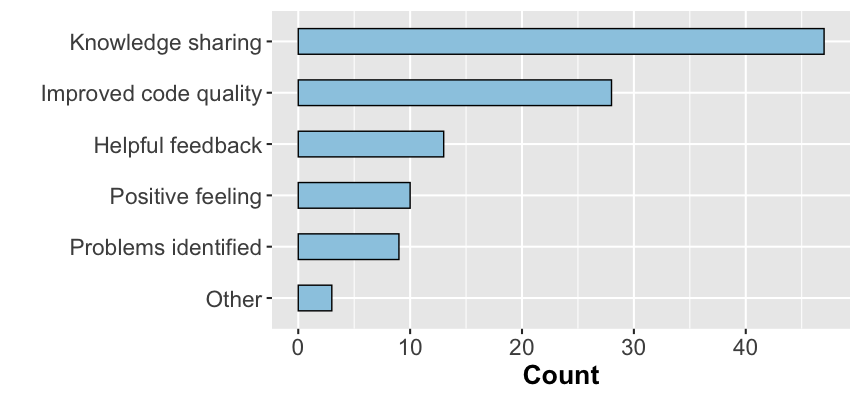}
	\caption{Positive experiences with peer code review}
	\label{fig_positive_experience}
\end{figure*}

\noindent
The responses to Q15 (Figure~\ref{fig_positive_experience}) identified five categories of positive experiences.
Relative to the most common positive response, \textbf{knowledge sharing}, one participant said \textit{``I find it [peer code review] to be a very cooperative process in which suggestions are welcomed and coders are looking for guidance. In a big project it is rare that anyone understands the whole picture. We rely on each others experience with their part of the project. It can lead to more complete understanding of the task.''}
Similarly another participant said \textit{``It [peer code review] leads to design discussions happening that would not have happened otherwise. It makes the team more knowledgeable about what work is happening on the project and how people are going about it.''}

The second most common positive experience was \textbf{improved code quality}.
One benefit to code quality was \textit{``people found mistakes in code that I wrote, that I would have missed and only found out about much further on the validation process.''}
peer code review also results in \textit{``much better code and a better understanding of different parts of the code.''}
This benefit is not limited to new project members because \textit{``everyone's code is better with peer review ... [including] the founders of the project who have 30+ years of programming experience.''}

\vspace{4pt}
\noindent
\textbf{Negative peer code review experiences}

\begin{figure*}[!htb]
	\includegraphics[width=0.75\textwidth]{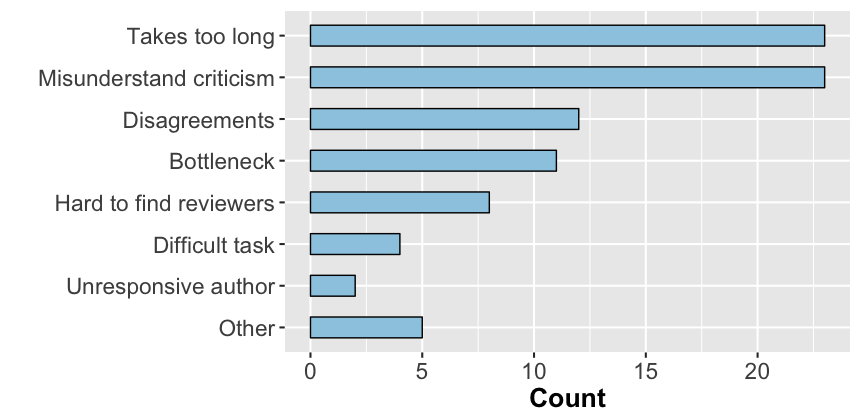}
	\caption{Negative experiences with peer code review}
	\label{fig_negative_experience}
\end{figure*}

\noindent
The responses to Q16 (Figure~\ref{fig_negative_experience}) resulted in seven categories of negative experiences.
The top two negative experiences are the peer code review process \textbf{takes too long} and code authors \textbf{misunderstand criticism}.

Regarding the fact that the peer code review process \textbf{takes too long}, one respondent said \textit{``it [peer code review] can be long and time consuming for very small changes, as the process must be followed for even a single character change if it affects results.''}
There are also problems when the \textit{``review process gets stalled while nit-picking irrelevant details''} or the \textit{``reviewer holds contribution hostage while asking to address unrelated problem at the same time.''}

Respondents are also concerned that the code authors will \textbf{misunderstand criticism}.
This concern makes reviewers less willing to provide criticism and concerned with how the criticism might affect team dynamics.
As one respondent noted \textit{``sometimes people get annoyed when they get feedback especially if they think they are experts. Sometimes people have misunderstandings about comments.''}

\subsection{\textit{\textbf{RQ2: What effect does peer code review have on research  software?}}}
\begin{figure*}[!htb]
	\includegraphics[width=0.75\textwidth]{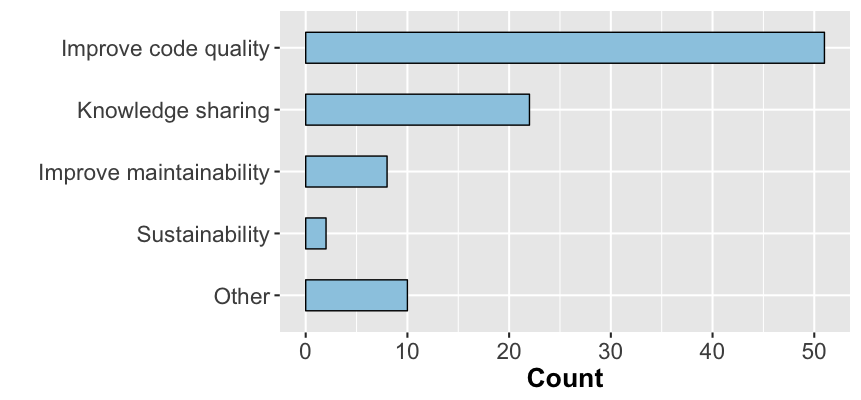}
	\caption{Explanation of why peer code review is important}
	\label{fig_code_review_is_important_agree}
\end{figure*}

The results of Q17 showed that 74 of the 84 participants \textbf{strongly agreed} that peer code review is important for their project.
Perhaps this result is not surprising given that these people chose to participate in a survey about peer code review practices.
In response to Q18, the participants gave four primary reasons for the positive responses (Figure~\ref{fig_code_review_is_important_agree}).
In terms of \textbf{improving code quality}, participants said \textit{``reviews are a way to improve the code and learn, without them bugs would proliferate and code quality would decrease''} and \textit{``it's a means of improving the quality of the software, and of improving the individuals who write it''} 
In terms of \textbf{knowledge sharing}, \textit{``peer code review helps ensure that at least two people have always looked at each piece of code, spreading out expertise. Additionally, it is a forum for learning from each other''}.

One of the few participants who did not think peer code review was important suggested that \textit{``requiring peer code review as a compliance activity is a waste of time. Since that is our project's policy, I view peer code review as a box to be checked''}.

\begin{figure*}[!htb]
	\includegraphics[width=0.75\textwidth]{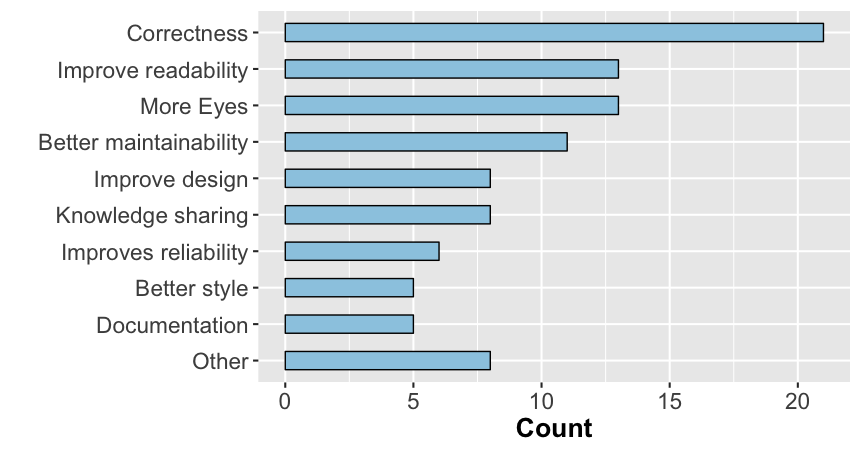}
	\caption{Explanation of why peer code review helps improve the code}
	\label{fig_code_review_improve_code_agree}
\end{figure*}

The results of Q19 showed that 77 out of the 84 participants \textbf{strongly agreed} that peer code review improves code.
In response to Q20, the participants gave nine reasons why they believed peer code review improved their code (Figure~\ref{fig_code_review_improve_code_agree}).
The most popular reason participants gave was that peer code review helped with \textbf{correctness}. 
As one respondent said \textit{``If you've written code yourself, it's hard to see the assumptions you've made. Others can spot these and ask you to clarify, also spot your mistakes.''}
Peer code review also improves code because it helps \textbf{improve readability} by \textit{``...make[ing] the codebase more uniform and improves the quality of the code"}.
In addition, having \textbf{more eyes} look at the code is beneficial because \textit{``having a second pair of eyes often catches issues the author missed or didn't consider''}.

\begin{figure*}[!htb]
	\includegraphics[width=0.75\textwidth]{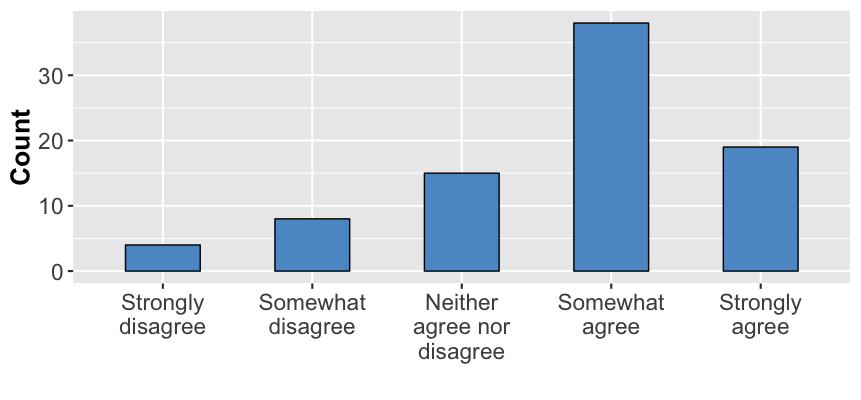}
	\caption{Peer code review helps decrease code complexity}
	\label{fig_decrease_code_complexity}
\end{figure*}

The responses to Q21 (Figure~\ref{fig_decrease_code_complexity}) shows that most participants agreed that peer code review helps decrease code complexity.
In response to Q22, participants explained their answer regarding the impact of peer code review on code complexity.
For the participants that thought peer code review helped reduce complexity, the explanations mirrored those in Figure~\ref{fig_code_review_improve_code_agree}. 
For example, peer code review \textit{``decreases complexity by solving problems using cleaner strategies, but may increase complexity in the near term by forcing it to handle corner cases that would not otherwise be discovered until later.''}

Conversely, some participants did not see peer code review as a means to reduce complexity.
One participant who did not see any relationship between the two said \textit{``peer code review often has no effect on code complexity whatsoever.  It is rare that someone skimming code for a few seconds will think of something drastic that the developer who spend days on it did not''}.
Another participant addressed the inherent complexity of some software saying \textit{``code sometimes needs to be performant; a less elegant solution that increases performance by $> 5\%$ is better for us than an elegant solution''}.

\subsection{\textit{\textbf{RQ3: What  difficulties  do  research  software  developers  face  with  code  review?}}}
The following text discusses the respondents perspectives on challenges and barriers in the peer code review process.

\vspace{4pt}
\noindent
\textbf{Challenges}

\begin{figure*}[!htb]
	\includegraphics[width=0.75\textwidth]{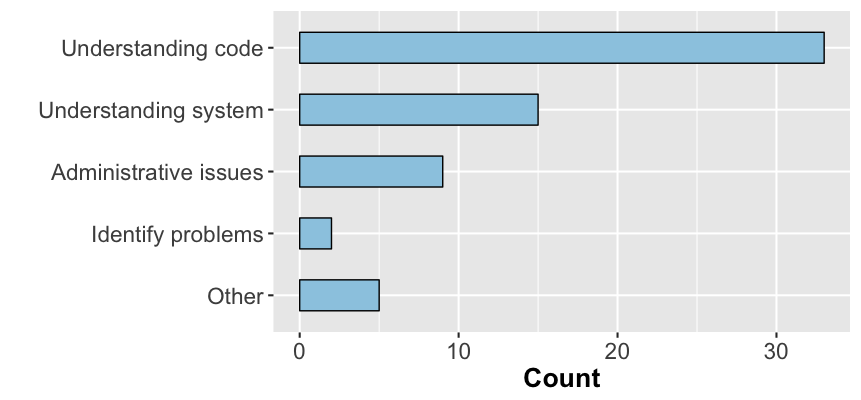}
	\caption{Challenging aspects of code review}
	\label{fig_challenges}
\end{figure*}

\noindent
The responses to Q23 (Figure~\ref{fig_challenges}) produced four types of challenges.
Overwhelmingly, the most common challenge is \textbf{understanding code}.
For some, \textit{``understanding code is significantly harder than writing code.''}
Understanding someone else's code requires time because the reviewer has \textit{``to work out exactly what added code is doing so that you can evaluate it.''}
In addition, reviewers have difficulty reading and understanding large changes.

The second most common challenge is \textbf{understanding the system}.
The reviewer needs a broader context than the specific change being reviewed.
The difficulty arises from \textit{``large diffs, diff is hard to read. Need good knowledge of existing code.''}
In addition, often \textit{``the code base is broad so the reviewer may know little about what the code is attempting to accomplish.''}

\vspace{4pt}
\noindent
\textbf{Barriers}

\begin{figure*}[!htb]
	\includegraphics[width=0.75\textwidth]{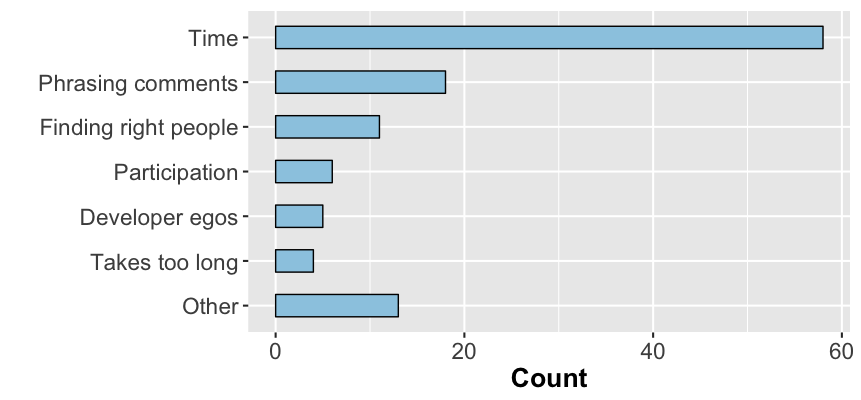}
	\caption{Barriers reviewers face when reviewing code}
	\label{fig_barriers}
\end{figure*}

\noindent
The responses to Q24 (Figure~\ref{fig_barriers}) produced six types of barriers.
Overwhelmingly, the most common barrier is \textbf{time}.
Peer code review takes \textit{``time away from the work you're `supposed' to be doing.''}
The context switch also costs time because \textit{``setup of the code under review also means making a commit of your local work and perhaps blowing away your own test data.''}

The second most common barrier relates to the \textbf{phrasing of comments}.
For example, \textit{``often our reviewers don't dare to criticise code, and need lots of encouragement to do so.  It sometimes helps to have pairs of reviewers instead of assigning them individually.''} 
Reviewers are also concerned about how their comments are perceived because \textit{``there is no tone in written comments (even though I could use emojis!) and I'm afraid the reviewee could be offended when I write `this is wrong!'  I try to write something more along the lines of `this should be x instead.'''}

Another barrier is \textbf{finding the right people}, that is \textit{``people with both domain knowledge and the coding knowledge.''}
It is difficult to find people who want to devote a lot of their time to reviewing code from other developers because
(1) There are not many people who possess both domain knowledge and software development knowledge and
(2) Researchers like to get recognition for their work~\citep{8588655,Eisty:2019:USP:3319008.3319351}, but do not receive any credit for performing peer code reviews. 

\subsection{\textit{\textbf{RQ4: What improvements to the peer code review process do research software developers need?}}}

\begin{figure*}[!htb]
	\includegraphics[width=0.75\textwidth]{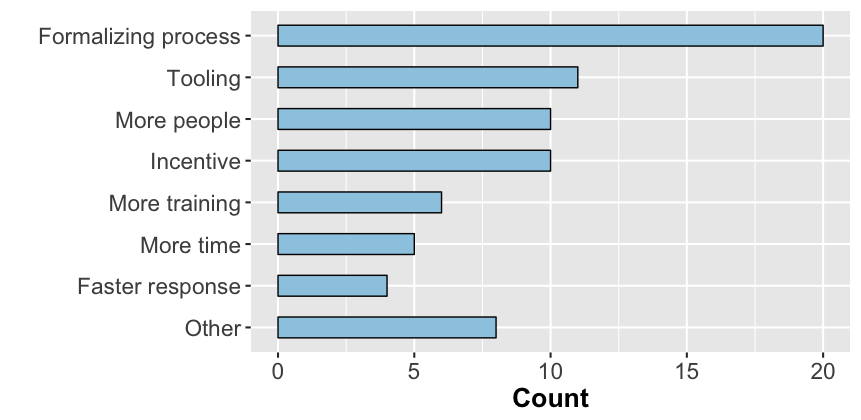}
	\caption{Improvement areas in the peer code review process}
	\label{fig_improve_code_review_process}
\end{figure*}

Because Q25 and Q26 produced similar responses, here we only discuss Q26.
The responses to Q26 (Figure~\ref{fig_improve_code_review_process}) produced in seven ways to improve the peer code review process.
The most common way, \textbf{formalizing process}, is likely due to the informal process followed by most research software developers.
One participant suggested using \textit{``a more formal structure of at least one science review followed by one technical review. It's currently a bit of a free-for-all.''}

Next \textit{automatic tooling} could relieve some of the burden on the reviewers by making \textit{``it easier to switch between forks and branches in operation.''}
Besides,\textit{``automated tools can save times for both review and reviewers''} which \textit{``include automatic analysis to release pressure from the reviewers.''}

Finally, there is a need for \textbf{more people}, \textbf{better incentives}, and \textbf{more training}.
One participant summed it up as \textit{``the main problem is the number of people actually doing it, and the amount of their time. Having said this: besides payment (often not possible) some other kind of reward might help with that.''}
Another indicated the need for \textit{``training more code reviewers so as to increase the reviewer base.''}

\section{Discussion}
\label{sec:Discussion}
In this section, we first summarize the key insights relative to each of the research questions, then discuss the implications of these results.
 
\subsection{Answers to Research Questions}
\noindent
\textbf{RQ1: How do research software developers perform peer code review?}

\noindent
Research software developers generally employ an informal peer code review process. 
This process identifies many types of problems ranging from typos to functionality.
Participants have both positive and negative experiences with peer code review.
Though they do not like when code authors misunderstand their criticism nor the time-consuming nature of the peer code review process, they do find the knowledge sharing and improved code quality to be positive aspects of the peer code review process. 

\vspace{8pt}
\noindent
\textbf{RQ2: What effect does peer code review have on research software?}

\noindent
Peer code review has an overall positive effect on research software. 
Participants found peer code review to be very important for their project.
However, the participants did not think peer code review had much effect on reducing code complexity because sometimes complex code is necessary to increase performance.
Overall, it is clear that peer code review greatly improves code quality by increasing correctness and clarity, which makes the code more readable and maintainable over time.

\vspace{8pt}
\noindent
\textbf{RQ3: What difficulties do research software developers face with peer code review?}

\noindent
There are many difficulties associated with the peer code review process in research software development. The most common difficulty reported by respondents is finding time to perform peer code reviews.
Another major challenge is understanding other people's code. 
In addition, sometimes understanding the whole system adds difficulty to the peer code review process.

\vspace{8pt}
\noindent
\textbf{RQ4: What improvements to the peer code review process do research software developers need?}

\noindent
There are several potential improvements to the peer code review process for research software.
The most important improvement is for projects to formalize the review process by including more people, more training, and providing compensation for performing peer code review.

\subsection{Implications of Results}
Overall, the findings from this study are similar to the literature on peer code review in traditional software engineering, as discussed in Section~\ref{sec:literature}.
Other than using a less structured process, research software developers achieve similar benefits and face similar difficulties when employing peer code review as do developers of commercial/IT or open-source software. 
While on one hand, these results may not be surprising, on the other hand, given the peculiarities of research software development, with its own constraints and issues, as highlighted in Section~\ref{sec:introduction}, these findings are encouraging and valuable.

Research software has often lagged behind commercial/IT and open-source software in adopting software engineering practices~\citep{5069156}.
This study shows that peer code review is as useful to research software developers as it is to developers of other kinds of software.
The results shows the various ways that peer code review can be beneficial for research software development.
This study also provides motivation for research software developers to adopt peer code review practices and for software engineering researchers to study the potential positive impacts other software engineering practices can have on research software development.

\section{Threats}
\label{sec:Threats}
This section describes the primary validity threats to this study.

\subsection{Internal Threats}
There are two primary threats to internal validity.
The first is whether participants understood the software engineering concepts in the same way we intended them.
Because the members of the target survey population are not traditional software engineers, it is possible that they lacked the necessary knowledge to properly answer the questions.
Because the participants used many appropriate software engineering terms to describe their practices, we believe they do have adequate software engineering knowledge to participate in this study.
Therefore, this threat is minimal.

The second threat to internal validity relates to the accuracy of the perceptions of the survey respondents.
Much of the data included in results is based upon the perceptions of the respondents.
Given that people often remember things differently than they actually occurred, it is possible that our results do not fully represent reality.
However, because much of the information necessary to answer our questions could only be gathered by direct interaction with participants (rather than through repository mining), we believe surveys are a valid approach.

\subsection{External Threats}
The results of this study may not be generalizable to all research software developers.
We attempted to recruit a wide sample for the survey, which is evident from the wide array of domains the participants' projects cover.
However, it is possible that the study participants are not representative of the population of research software developers.
As we collected perceptions of the developers, it could vary from person to person. 
Even people working on the same project could potentially have different perceptions and experience about the process.

For the interviews, we drew from two closed populations (projects and NCSA and participants in a conference for a large project).
For the surveys, the respondents were those who were on one or more mailing lists related to the development of research software, therefore suggesting they may have more than casual interest in the topic.
Furthermore, while it is clear that the participants are all research software developers, some of the responses suggest that they may be more interested in peer code review than the average research software developer.
In addition, they took time to answer a survey about peer code review.
Therefore, the responses may be biased towards those developers who are already predisposed towards the use of peer code review.
Such a skew in the population would produce results that may be less descriptive of, although no less relevant to, the larger research software developer community.

\subsection{Construct Threats}
The primary construct validity threat is the participants may misunderstand the questions. 
We took great care in writing the survey questions and verified them by expert research software developers and software engineering researchers.
In addition, we explained the questions to the interview participants without biasing them so they could used their own judgment to respond. 

\section{Conclusion and Future Work}
\label{sec:Conclusion}
In this paper, we report insights about peer code review in research software based upon the responses from 84 survey and interview participants. 
We also report the effects of the code review process on their projects, difficulties they face in performing code review, and recommendations to improve the process.
Considering that our survey targeted a specific type of software projects, those focused on research software, and previous experience shows that research software projects are often less advanced in the software engineering process, we were still able to attract a good number of survey respondents.

Our results show the importance and usefulness of peer code review as an essential and important practice to help research software developers produce high-quality, trustworthy software. 
However, peer code review does introduce some challenges related to understanding the need for it, communication misunderstandings that occur during the process, necessity of tools, and available time. 
In some cases peer code review may seem like an extra burden to the developers and increase the time required for development.
Though peer code review presents some challenges and barriers to being established as a core practice in research software setting, it can improve research software in many ways and overcome challenges that cannot be addressed by testing or other software quality practices. 

Building on the findings of this paper, our future plan is to work more closely with research software projects to better understand how these results apply in practice.
By interacting directly with research software developers as they write software, we can conduct direct observational studies to gather information in real-time rather than relying on the type of retrospective reports that we obtained from our interviews and surveys.
The results from these direct case studies will help to expand the findings of the interviews and surveys and provide additional useful and practical insights for research software developers.

We plan to make these results actionable by raising awareness within the research software community about the benefits and importance of peer code review in their projects. 
The findings reported in this paper provide evidence that adopting peer code review can improve the quality of research software.
We have begun to to provide training on contemporary peer code review process to research software developers and plan to continue providing that training in the future.
The training on peer code review includes results from this paper, descriptions of available methods for peer code review, and best practices in conduct peer code review (both for the code author and for the reviewer).
These training events help research software developers utilize peer code review in their projects and ultimately build high quality research software and produce trustworthy results.

\begin{acknowledgements}
The authors acknowledge partial support from the US National Science Foundation (1445344).
Nasir Eisty would like to thank his sponsors for the NCSA summer internship, Drs. Gabrielle Allen, Roland Hass and Daniel S. Katz.
We would also like to thank the interview and survey participants.
\end{acknowledgements}

\bibliographystyle{spbasic}       
\bibliography{references}   

\begin{thebibliography}{36}
\providecommand{\natexlab}[1]{#1}
\providecommand{\url}[1]{{#1}}
\providecommand{\urlprefix}{URL }
\expandafter\ifx\csname urlstyle\endcsname\relax
  \providecommand{\doi}[1]{DOI~\discretionary{}{}{}#1}\else
  \providecommand{\doi}{DOI~\discretionary{}{}{}\begingroup
  \urlstyle{rm}\Url}\fi
\providecommand{\eprint}[2][]{\url{#2}}

\bibitem[{Ackerman et~al.(1989)Ackerman, Buchwald, and Lewski}]{28121}
Ackerman AF, Buchwald LS, Lewski FH (1989) Software inspections: an effective
  verification process. IEEE Software 6(3):31--36, \doi{10.1109/52.28121}

\bibitem[{Bacchelli and Bird(2013)}]{Bacchelli:2013:EOC:2486788.2486882}
Bacchelli A, Bird C (2013) Expectations, outcomes, and challenges of modern
  code review. In: Proceedings of the 2013 International Conference on Software
  Engineering, IEEE Press, Piscataway, NJ, USA, ICSE '13, pp 712--721,
  \urlprefix\url{http://dl.acm.org/citation.cfm?id=2486788.2486882}

\bibitem[{Beller et~al.(2014{\natexlab{a}})Beller, Bacchelli, Zaidman, and
  Juergens}]{10.1145/2597073.2597082}
Beller M, Bacchelli A, Zaidman A, Juergens E (2014{\natexlab{a}}) Modern code
  reviews in open-source projects: Which problems do they fix? In: Proceedings
  of the 11th Working Conference on Mining Software Repositories, Association
  for Computing Machinery, New York, NY, USA, MSR 2014, p 202–211,
  \urlprefix\url{https://doi.org/10.1145/2597073.2597082}

\bibitem[{Beller et~al.(2014{\natexlab{b}})Beller, Bacchelli, Zaidman, and
  Juergens}]{Beller:2014:MCR:2597073.2597082}
Beller M, Bacchelli A, Zaidman A, Juergens E (2014{\natexlab{b}}) Modern code
  reviews in open-source projects: Which problems do they fix? In: Proceedings
  of the 11th Working Conference on Mining Software Repositories, ACM, New
  York, NY, USA, MSR 2014, pp 202--211, \doi{10.1145/2597073.2597082},
  \urlprefix\url{http://doi.acm.org/10.1145/2597073.2597082}

\bibitem[{Bosu and Carver(2013)}]{6681346}
Bosu A, Carver JC (2013) Impact of peer code review on peer impression
  formation: A survey. In: 2013 ACM / IEEE International Symposium on Empirical
  Software Engineering and Measurement, pp 133--142, \doi{10.1109/ESEM.2013.23}

\bibitem[{Bosu et~al.(2014)Bosu, Carver, Guadagno, Bassett, McCallum, and
  Hochstein}]{BOSU20144}
Bosu A, Carver J, Guadagno R, Bassett B, McCallum D, Hochstein L (2014) Peer
  impressions in open source organizations: A survey. Journal of Systems and
  Software 94:4 -- 15, \doi{https://doi.org/10.1016/j.jss.2014.03.061},
  \urlprefix\url{http://www.sciencedirect.com/science/article/pii/S0164121214000818}

\bibitem[{Bosu et~al.(2015)Bosu, Greiler, and Bird}]{7180075}
Bosu A, Greiler M, Bird C (2015) Characteristics of useful code reviews: An
  empirical study at microsoft. In: 2015 IEEE/ACM 12th Working Conference on
  Mining Software Repositories, pp 146--156, \doi{10.1109/MSR.2015.21}

\bibitem[{Bosu et~al.(2017)Bosu, Carver, Bird, Orbeck, and Chockley}]{7484733}
Bosu A, Carver JC, Bird C, Orbeck J, Chockley C (2017) Process aspects and
  social dynamics of contemporary code review: Insights from open source
  development and industrial practice at microsoft. IEEE Transactions on
  Software Engineering 43(1):56--75

\bibitem[{Carver et~al.(2016)Carver, Chue~Hong, and Thiruvathukal}]{1958562}
Carver J, Chue~Hong N, Thiruvathukal G (2016) Software Engineering for Science.
  CRC Press

\bibitem[{Carver(2008)}]{10.1145/1370175.1370252}
Carver JC (2008) Se-cse 2008: The first international workshop on software
  engineering for computational science and engineering. In: Companion of the
  30th International Conference on Software Engineering, Association for
  Computing Machinery, New York, NY, USA, ICSE Companion ’08, p 1071–1072,
  \doi{10.1145/1370175.1370252},
  \urlprefix\url{https://doi.org/10.1145/1370175.1370252}

\bibitem[{Carver and Eisty(2021)}]{dataset}
Carver JC, Eisty N (2021) Data set from survey on peer code review in research
  software. \doi{10.6084/m9.figshare.14736468},
  \urlprefix\url{https://figshare.com/articles/dataset/_/14736468/0}

\bibitem[{{Ciolkowski} et~al.(2003){Ciolkowski}, {Laitenberger}, and
  {Biffl}}]{1241366}
{Ciolkowski} M, {Laitenberger} O, {Biffl} S (2003) Software reviews, the state
  of the practice. IEEE Software 20(6):46--51, \doi{10.1109/MS.2003.1241366}

\bibitem[{{Clune} and {Rood}(2011)}]{5999647}
{Clune} T, {Rood} R (2011) Software testing and verification in climate model
  development. IEEE Software 28(6):49--55, \doi{10.1109/MS.2011.117}

\bibitem[{Eisty et~al.(2018)Eisty, Thiruvathukal, and Carver}]{8588655}
Eisty NU, Thiruvathukal GK, Carver JC (2018) A survey of software metric use in
  research software development. In: 2018 IEEE 14th International Conference on
  e-Science (e-Science), pp 212--222

\bibitem[{Eisty et~al.(2019)Eisty, Thiruvathukal, and
  Carver}]{Eisty:2019:USP:3319008.3319351}
Eisty NU, Thiruvathukal GK, Carver JC (2019) Use of software process in
  research software development: A survey. In: Proceedings of the Evaluation
  and Assessment on Software Engineering, ACM, New York, NY, USA, EASE '19, pp
  276--282, \doi{10.1145/3319008.3319351},
  \urlprefix\url{http://doi.acm.org/10.1145/3319008.3319351}

\bibitem[{{Fagan}(1976)}]{5388086}
{Fagan} ME (1976) Design and code inspections to reduce errors in program
  development. IBM Systems Journal 15(3):182--211, \doi{10.1147/sj.153.0182}

\bibitem[{{Faulk} et~al.(2009){Faulk}, {Loh}, {Vanter}, {Squires}, and
  {Votta}}]{5337642}
{Faulk} S, {Loh} E, {Vanter} MLVD, {Squires} S, {Votta} LG (2009) Scientific
  computing's productivity gridlock: How software engineering can help.
  Computing in Science Engineering 11(6):30--39

\bibitem[{{Hannay} et~al.(2009){Hannay}, {MacLeod}, {Singer}, {Langtangen},
  {Pfahl}, and {Wilson}}]{5069155}
{Hannay} JE, {MacLeod} C, {Singer} J, {Langtangen} HP, {Pfahl} D, {Wilson} G
  (2009) How do scientists develop and use scientific software? In: 2009 ICSE
  Workshop on Software Engineering for Computational Science and Engineering,
  pp 1--8, \doi{10.1109/SECSE.2009.5069155}

\bibitem[{Heaton and Carver(2015)}]{HEATON2015207}
Heaton D, Carver JC (2015) Claims about the use of software engineering
  practices in science: A systematic literature review. Information and
  Software Technology 67:207 -- 219,
  \doi{https://doi.org/10.1016/j.infsof.2015.07.011},
  \urlprefix\url{http://www.sciencedirect.com/science/article/pii/S0950584915001342}

\bibitem[{{Hook} and {Kelly}(2009)}]{5069163}
{Hook} D, {Kelly} D (2009) Testing for trustworthiness in scientific software.
  In: 2009 ICSE Workshop on Soft. Eng. for Computational Science and Eng., pp
  59--64, \doi{10.1109/SECSE.2009.5069163}

\bibitem[{Kanewala and Bieman(2014)}]{KANEWALA20141219}
Kanewala U, Bieman JM (2014) Testing scientific software: A systematic
  literature review. Information and Software Technology 56(10):1219 -- 1232,
  \doi{https://doi.org/10.1016/j.infsof.2014.05.006},
  \urlprefix\url{http://www.sciencedirect.com/science/article/pii/S0950584914001232}

\bibitem[{Mantyla and Lassenius(2009)}]{Mantyla:2009:TDR:1591905.1592371}
Mantyla MV, Lassenius C (2009) What types of defects are really discovered in
  code reviews? IEEE Trans Softw Eng 35(3):430--448, \doi{10.1109/TSE.2008.71},
  \urlprefix\url{http://dx.doi.org/10.1109/TSE.2008.71}

\bibitem[{McIntosh et~al.(2014)McIntosh, Kamei, Adams, and
  Hassan}]{10.1145/2597073.2597076}
McIntosh S, Kamei Y, Adams B, Hassan AE (2014) The impact of code review
  coverage and code review participation on software quality: A case study of
  the qt, vtk, and itk projects. In: Proceedings of the 11th Working Conference
  on Mining Software Repositories, Association for Computing Machinery, New
  York, NY, USA, MSR 2014, p 192–201, \doi{10.1145/2597073.2597076},
  \urlprefix\url{https://doi.org/10.1145/2597073.2597076}

\bibitem[{Morales et~al.(2015)Morales, McIntosh, and Khomh}]{7081827}
Morales R, McIntosh S, Khomh F (2015) Do code review practices impact design
  quality? a case study of the qt, vtk, and itk projects. In: 2015 IEEE 22nd
  International Conference on Software Analysis, Evolution, and Reengineering
  (SANER), pp 171--180, \doi{10.1109/SANER.2015.7081827}

\bibitem[{Nangia and Katz(2017)}]{nangia_katz_2017}
Nangia U, Katz DS (2017) Track 1 paper: Surveying the u.s. national
  postdoctoral association regarding software use and training in research.
  \doi{10.6084/m9.figshare.5328442.v3},
  \urlprefix\url{https://figshare.com/articles/Track_1_Paper_Surveying_the_U_S_National_Postdoctoral_Association_Regarding_Software_Use_and_Training_in_Research/5328442/3}

\bibitem[{Petre and Wilson(2014)}]{article159}
Petre M, Wilson G (2014) Code review for and by scientists. In: Proc. 2nd
  Workshop on Sustainable Software for Science: Practice and Experience

\bibitem[{Potvin and Levenberg(2016)}]{Potvin:2016:WGS:2963119.2854146}
Potvin R, Levenberg J (2016) Why google stores billions of lines of code in a
  single repository. Commun ACM 59(7):78--87, \doi{10.1145/2854146},
  \urlprefix\url{http://doi.acm.org/10.1145/2854146}

\bibitem[{{Remmel} et~al.(2012){Remmel}, {Paech}, {Bastian}, and
  {Engwer}}]{6086527}
{Remmel} H, {Paech} B, {Bastian} P, {Engwer} C (2012) System testing a
  scientific framework using a regression-test environment. Computing in
  Science Engineering 14(2):38--45

\bibitem[{Rigby et~al.(2012)Rigby, Cleary, Painchaud, Storey, and
  German}]{6148202}
Rigby P, Cleary B, Painchaud F, Storey M, German D (2012) Contemporary peer
  review in action: Lessons from open source development. IEEE Software
  29(6):56--61, \doi{10.1109/MS.2012.24}

\bibitem[{Rigby et~al.(2008)Rigby, German, and
  Storey}]{Rigby:2008:OSS:1368088.1368162}
Rigby PC, German DM, Storey MA (2008) Open source software peer review
  practices: A case study of the apache server. In: Proceedings of the 30th
  International Conference on Software Engineering, ACM, New York, NY, USA,
  ICSE '08, pp 541--550, \doi{10.1145/1368088.1368162},
  \urlprefix\url{http://doi.acm.org/10.1145/1368088.1368162}

\bibitem[{Sadowski et~al.(2018)Sadowski, S\"{o}derberg, Church, Sipko, and
  Bacchelli}]{Sadowski:2018:MCR:3183519.3183525}
Sadowski C, S\"{o}derberg E, Church L, Sipko M, Bacchelli A (2018) Modern code
  review: A case study at google. In: Proceedings of the 40th International
  Conference on Software Engineering: Software Engineering in Practice, ACM,
  New York, NY, USA, ICSE-SEIP '18, pp 181--190, \doi{10.1145/3183519.3183525},
  \urlprefix\url{http://doi.acm.org/10.1145/3183519.3183525}

\bibitem[{{Sanders} and {Kelly}(2008)}]{4548404}
{Sanders} R, {Kelly} D (2008) Dealing with risk in scientific software
  development. IEEE Software 25(4):21--28

\bibitem[{Segal(2005)}]{article_segal}
Segal J (2005) When software engineers met research scientists: A case study.
  Empirical Software Engineering 10, \doi{10.1007/s10664-005-3865-y}

\bibitem[{{Segal}(2009)}]{5069156}
{Segal} J (2009) Some challenges facing software engineers developing software
  for scientists. In: 2009 ICSE Workshop on Software Engineering for
  Computational Science and Engineering, pp 9--14

\bibitem[{Sutherland and Venolia(2009)}]{5070996}
Sutherland A, Venolia G (2009) Can peer code reviews be exploited for later
  information needs? In: 2009 31st International Conference on Software
  Engineering - Companion Volume, pp 259--262,
  \doi{10.1109/ICSE-COMPANION.2009.5070996}

\bibitem[{Vilkomir et~al.(2008)Vilkomir, Swain, Poore, and
  Clarno}]{10.1007/978-3-540-69389-5_34}
Vilkomir SA, Swain WT, Poore JH, Clarno KT (2008) Modeling input space for
  testing scientific computational software: A case study. In: Bubak M, van
  Albada GD, Dongarra J, Sloot PMA (eds) Computational Science -- ICCS 2008,
  Springer Berlin Heidelberg, pp 291--300

\end{thebibliography}

\end{document}